\documentclass[conference]{IEEEtran}
\IEEEoverridecommandlockouts

\usepackage{
amsmath,
graphicx,
cite,
multirow,
booktabs,
amssymb,
enumitem,
url,
amsfonts,
algorithmic,
textcomp,
xcolor,
hyperref,
}

\def\BibTeX{{\rm B\kern-.05em{\sc i\kern-.025em b}\kern-.08em
    T\kern-.1667em\lower.7ex\hbox{E}\kern-.125emX}}
    
\begin{document}

\title{QAMRO: Quality-aware Adaptive Margin Ranking Optimization for Human-aligned Assessment of Audio Generation Systems}

\author{\IEEEauthorblockN{
Chien-Chun Wang\IEEEauthorrefmark{1},
Kuan-Tang Huang\IEEEauthorrefmark{1},
Cheng-Yeh Yang\IEEEauthorrefmark{1},
Hung-Shin Lee\IEEEauthorrefmark{3},
Hsin-Min Wang\IEEEauthorrefmark{2},
and Berlin Chen\IEEEauthorrefmark{1}
}
\IEEEauthorblockA{\IEEEauthorrefmark{1}Dept. Computer Science and Information Engineering, National Taiwan Normal University, Taiwan}
\IEEEauthorblockA{\IEEEauthorrefmark{2}Institute of Computer Science, Academia Sinica, Taiwan}
\IEEEauthorblockA{\IEEEauthorrefmark{3}United Link Co., Ltd., Taiwan}
}

\maketitle

\begin{abstract}

Evaluating audio generation systems, including text-to-music (TTM), text-to-speech (TTS), and text-to-audio (TTA), remains challenging due to the subjective and multi-dimensional nature of human perception.
Existing methods treat mean opinion score (MOS) prediction as a regression problem, but standard regression losses overlook the relativity of perceptual judgments.
To address this limitation, we introduce QAMRO, a novel Quality-aware Adaptive Margin Ranking Optimization framework that seamlessly integrates regression objectives from different perspectives, aiming to highlight perceptual differences and prioritize accurate ratings.
Our framework leverages pre-trained audio-text models such as CLAP and Audiobox-Aesthetics, and is trained exclusively on the official AudioMOS Challenge 2025 dataset.
It demonstrates superior alignment with human evaluations across all dimensions, significantly outperforming robust baseline models.

\end{abstract}


\section{Introduction}

Audio generation systems, including text-to-music (TTM) \cite{agostinelli2023,dong2023}, text-to-speech (TTS) \cite{ju2024,chen2024,chen2025}, and text-to-audio (TTA) \cite{huang2023a,borsos2023}, have made rapid progress driven by breakthroughs such as diffusion models and large language models \cite{liu2023,copet2023,huang2023,ji2023,zhu2023,lu2023,schneider2023,ziv2024}.
These systems can generate expressive audio from text, supporting applications in areas such as music production, content creation, and interactive media.
A key challenge is to evaluate the perceptual quality of the generated audio.
While subjective mean opinion scores (MOS) can provide reliable assessments, they are expensive, non-scalable, and difficult to reproduce.
In contrast, objective metrics, such as Fréchet Audio Distance (FAD) \cite{kilgour2019} and Inception Score (IS) \cite{barratt2018}, are often inconsistent with human preferences, especially in terms of semantic alignment and perceptual quality.

As a result, MOS prediction has been formulated as a regression task, where the model estimates human perceptual quality scores based on audio and optionally text \cite{liu2025}.
Most previous research efforts \cite{tseng2021,saeki2022,reddy2022,fu2023,ariyanti2025} employ training losses such as Mean Absolute Error (MAE) or Mean Squared Error (MSE) to minimize the prediction errors of the models.
However, these methods do not account for the relative rankings among samples, a critical component of perceptual assessment.
Absolute MOS values may vary across datasets and annotators, while relative preferences are often more consistent and meaningful in practical use cases such as system comparison and model selection.
Although relatively underexplored in MOS prediction, the development of effective ranking loss functions has recently gained traction in several application domains \cite{menon2012,engilberge2019,wang2019,qian2020,oksuz2020,swezey2021,bruch2021,oksuz2021,wu2021,liu2021,zhu2022,sheng2023,hu2023,xiao2023,yavuz2024,durmus2024,tan2025,peng2025} due to their ability to better capture relative rankings in subjective regression tasks.
However, standard ranking loss functions typically adopt a fixed margin and treat all sample pairs uniformly, ignoring perceptual differences and the different importance of high-quality content.

In light of these limitations, we propose a novel modeling framework to introduce a ranking-based perspective to the MOS prediction task.
Specifically, we present a \textbf{Q}uality-aware \textbf{A}daptive \textbf{M}argin \textbf{R}anking \textbf{O}ptimization (QAMRO) strategy that enhances the training of MOS prediction models by encouraging correct pairwise rankings.
Unlike conventional ranking losses \cite{burges2006,cao2007,burges2010}, QAMRO adjusts the margin according to perceptual score gaps and gives more weight to high-quality samples.
We conducted experiments strictly on the official AudioMOS Challenge 2025 datasets (i.e., in a closed-set setting), and the results show that incorporating ranking-aware supervision can improve alignment with human judgment while maintaining accurate absolute score predictions.
This demonstrates the effectiveness of our strategy in perceptual regression, thereby suggesting a promising direction for future audio assessment research.

\section{Proposed Method}

\subsection{Quality-aware Adaptive Margin Ranking Optimization}

To better align model predictions with human perceptual preferences, we design a quality-aware adaptive margin ranking loss, which puts focus on preserving the relative ranking of predicted scores, complementarily to standard regression losses that normally minimize absolute prediction errors.

Given a set of sample pairs $\mathcal{P} = \{(i, j) \mid y_i \neq y_j\}$ in a mini-batch, where $y_i$ and $y_j$ denote the ground-truth MOS scores of samples $i$ and $j$, respectively, we define two ranking-based loss functions.
The first is the \textit{margin ranking loss}, which is a standard ranking loss that enforces a fixed margin $m$ between the predicted scores of samples with different ratings, with all sample pairs being equally weighted:
\vspace{-5pt}
\begin{equation}\label{eq:mr}
\begin{split}
\mathcal{L}_{\text{MR}} = \frac{1}{|\mathcal{P}|} \sum_{(i,j) \in \mathcal{P}} \max\Big(0,\ -\operatorname{sign}(y_i - y_j) \cdot (\hat{y}_i - \hat{y}_j) + m \Big),
\end{split}
\end{equation}
where $\hat{y}_i$ and $\hat{y}_j$ mark, respectively, the predicted scores for samples $i$ and $j$, and $\operatorname{sign}(x)$ equals $+1$ if $x > 0$, and $-1$ otherwise, indicating the correct ranking.
The second is the proposed \textit{quality-aware adaptive margin ranking loss}, which is defined by
\vspace{-5pt}
\begin{equation}\label{eq:qamro}
\begin{split}
\mathcal{L}_{\text{QAMRO}} = \frac{1}{|\mathcal{P}|} \sum_{(i,j) \in \mathcal{P}} q_{ij} & \cdot \max\Big(0,\ -\operatorname{sign}(y_i - y_j) \\
& \cdot (\hat{y}_i - \hat{y}_j) + \alpha \cdot |y_i - y_j| \Big),
\end{split}
\end{equation}
where $q_{ij}$ is a quality-aware weight, and $\alpha$ denotes a margin scaling factor.
This formulation remedies the deficiency of the conventional pairwise ranking loss through two novel countermeasures.
First, it employs a data-dependent margin (\textit{cf.} the term $\alpha \cdot |y_i - y_j|$ in Eq.~\ref{eq:qamro}) that varies based on the absolute difference of the ground-truth MOS values, thereby capturing perceptual discrepancies.
Second, a quality-aware weighting mechanism is incorporated:
\vspace{-5pt}
\begin{equation}\label{eq:weight}
q_{ij} = 1 + (\beta - 1) \cdot \max(\tilde{y}_i, \tilde{y}_j),
\end{equation}
where $\tilde{y}_i$ and $\tilde{y}_j \in [0,1]$ are the normalized ground-truth MOS scores, and the preference factor $\beta > 1$ modulates the strength of the weights. 
This accentuates sample pairs that contain at least one high-quality utterance, thus encouraging the model to rank such cases more reliably.

In summary, $\mathcal{L}_{\text{MR}}$ serves as a baseline with a uniform margin and equal weighting, assuming a fixed perceptual threshold for distinguishing the score ranking.
As a soft-margin formulation, it allows for subtle violations of the target ordering while penalizing larger discrepancies.
Our $\mathcal{L}_{\text{QAMRO}}$ improves on this by adapting the margin according to the perceptual gap $|y_i - y_j|$ and giving higher weights to high-quality samples via the weighting term $q_{ij}$.
This reflects the intuition that human perceptual differences vary with the score distance and that errors in high-quality regions are usually more critical.

\subsection{MOS Prediction Tasks Across Audio Generation Systems}

To demonstrate the generalizability and effectiveness of our proposed optimization strategy, we first apply it to a MOS prediction model for TTM systems.
The model is built upon the pre-trained CLAP model \cite{wu2023}, which captures audio-text joint semantics.
The music clips are encoded using the CLAP audio encoder, while the textual prompts are embedded via the CLAP text encoder.
Two separate three-layer multi-layer perceptrons (MLPs) are employed: one regresses the musical impression score from the audio-only embeddings, and the other estimates the text alignment score using the concatenated audio and text embeddings.
The training objective combines our proposed QAMRO loss with the Huber loss \cite{huber1992}. The QAMRO loss enforces consistency with human perceptual rankings, especially in high-quality regions, while the Huber loss enhances robustness by reducing the impact of outliers.

Based on the success of this strategy for TTM, we extend our framework to a more general MOS prediction setting, including TTS, TTA, and TTM.
We adopt the pre-trained Audiobox-Aesthetics model \cite{tjandra2025} as the backbone.
Audio input streams are processed through a convolutional encoder and a Transformer encoder to extract deep audio representations.
These embeddings are fed into four independent MLP prediction heads, each for one evaluation dimension: production quality, production complexity, content enjoyment, and content usefulness.
The model is trained with a multi-objective loss function combining our QAMRO loss with traditional regression objectives, consequently balancing perceptual alignment and score accuracy for different content types.

\section{Experimental Setups}

\subsection{Datasets}

Our experiments were conducted solely on the AudioMOS Challenge 2025.
For the TTM MOS prediction task, we used the MusicEval dataset \cite{liu2025}, which comprises 2,748 music clips generated by 31 text-to-music systems, each conditioned on one of 384 unique text prompts.
Each clip was rated by five annotators selected from a pool of 14 music experts, along two perceptual dimensions: musical impression and alignment with the given text.
To further evaluate the generalizability of our framework across various types of audio content, we employed the AES-Natural dataset \cite{tjandra2025}, which contains 2,950 audio samples covering speech, music, and general audio content.
Specifically, it includes 950 speech samples (from EARS, LibriTTS, and Common Voice 13.0), 1,000 music samples (from MUSDB18-HQ and MusicCaps), and 1,000 general audio samples from AudioSet.
Each sample was annotated by 10 experts with backgrounds in audio or music.
To support multi-faceted evaluation, each clip was rated along four perceptual axes: production quality, production complexity, content enjoyment, and content usefulness.

\subsection{Configurations}

All models were trained with a batch size of 256 using the SGD optimizer \cite{ruder2017} and a fixed learning rate of 0.0005.
For the proposed ranking loss, we set $\alpha=0.2$ and $\beta=7.0$ in Eqs.~\ref{eq:qamro} and \ref{eq:weight}, respectively.
Early stopping was applied if the validation loss did not improve for 20 epochs.
Our code is available at \url{https://github.com/JethroWangSir/QAMRO}.

\begin{table*}[t]
\small
\caption{
Results and ablations on the MusicEval dataset, evaluating musical impression and textual alignment.
Note that ``w/o Adaptive Margin'' denotes replacing the proposed margin-based ranking loss $\mathcal{L}_{\text{QAMRO}}$ with a fixed-margin ranking loss $\mathcal{L}_{\text{MR}}$.
}
\vspace{-8pt}
\label{tab:track1}
\centering
\setlength{\tabcolsep}{9pt}
\begin{tabular}{lcccccccc}
\toprule
\multirow{2}{*}{\bf{Model}} & \multicolumn{4}{c}{\bf{Musical Impression}} & \multicolumn{4}{c}{\bf{Textual Alignment}} \\ \cmidrule(lr){2-5} \cmidrule(lr){6-9}
 & \bf{MSE~$\downarrow$} & \bf{LCC~$\uparrow$} & \bf{SRCC~$\uparrow$} & \bf{KTAU~$\uparrow$} & \bf{MSE~$\downarrow$} & \bf{LCC~$\uparrow$} & \bf{SRCC~$\uparrow$} & \bf{KTAU~$\uparrow$} \\
\toprule
MusicEval-baseline \cite{liu2025} & 0.247 & 0.851 & 0.845 & 0.655 & 0.143 & 0.819 & 0.779 & 0.578 \\
\bf{QAMRO} & \bf{0.139} & \bf{0.961} & \bf{0.972} & \bf{0.876} & 0.109 & \bf{0.918} & \bf{0.916} & \bf{0.763} \\
\midrule
w/o Quality-aware Weighting & 0.172 & 0.910 & 0.924 & 0.779 & \bf{0.101} & 0.894 & 0.885 & 0.707 \\
~~w/o Adaptive Margin & 0.214 & 0.886 & 0.875 & 0.687 & 0.134 & 0.875 & 0.838 & 0.643 \\
\bottomrule
\end{tabular}
\vspace{-14pt}
\end{table*}

\subsection{Evaluation Metrics}

We evaluated system performance using several objective system-level metrics: MSE, linear correlation coefficient (LCC), Spearman rank correlation coefficient (SRCC), and Kendall Tau rank correlation (KTAU) \cite{cooper2022}.
These metrics are computed by first averaging MOS predictions across all clips generated by each system, then comparing these system-level predictions with corresponding human-annotated system-level MOS scores.
System-level evaluation was adopted because, in real-world scenarios, assessing and ranking complete speech systems offers greater practical relevance than evaluating individual clips.
MSE ranges from 0 to $\infty$, with lower values indicating better performance.
LCC, SRCC, and KTAU range from $-1$ to $1$, where higher values indicate stronger correlation and better alignment with human judgments.
SRCC, in particular, reflects ranking consistency across systems.

\begin{table}[t]
\small
\caption{Results and ablations on the AES-Natural dataset, reporting system-level SRCC for PQ, PC, CE, and CU.}
\vspace{-8pt}
\label{tab:track2}
\centering
\setlength{\tabcolsep}{5.5pt}
\begin{tabular}{lcccc}
\toprule
\bf{Model} & \bf{PQ} & \bf{PC} & \bf{CE} & \bf{CU} \\
\toprule
Audiobox-Aesthetics \cite{tjandra2025} & 0.864 & 0.938 & 0.845 & 0.809 \\
\bf{QAMRO} & \bf{0.883} & \bf{0.942} & \bf{0.869} & \bf{0.852} \\
\midrule
w/o Quality-aware Weighting & 0.874 & 0.940 & 0.861 & 0.843 \\
~~w/o Adaptive Margin & 0.871 & 0.939 & 0.853 & 0.829 \\
\bottomrule
\end{tabular}
\vspace{-10pt}
\end{table}

\section{Results and Discussion}

\subsection{Performance Analysis on MusicEval}

Table~\ref{tab:track1} reports the results and ablations of our proposed framework, QAMRO, on the MusicEval dataset.
QAMRO achieves the highest overall performance, with SRCC scores of 0.972 for musical impression and 0.916 for textual alignment, outperforming the MusicEval-baseline \cite{liu2025}.
For a fair comparison, both our framework and the baseline used the Huber loss for score regression.
These results demonstrate the effectiveness of our quality-aware adaptive margin ranking optimization strategy.
Although SRCC is most directly influenced by our ranking loss, our framework also improves other metrics like MSE, LCC, and KTAU.
This shows that it improves regression quality without sacrificing it, strengthening overall learning rather than focusing solely on ranking.

Ablation studies further validate the contribution of each component.
Removing the quality-aware weighting leads to a noticeable drop in performance for both musical impression and textual alignment.
Further excluding the adaptive margin mechanism results in an even larger degradation.
These results demonstrate the importance of both the quality-aware weighting and adaptive margin for optimal ranking performance.

\begin{figure}[t]
\centering
\includegraphics[width=0.98\linewidth]{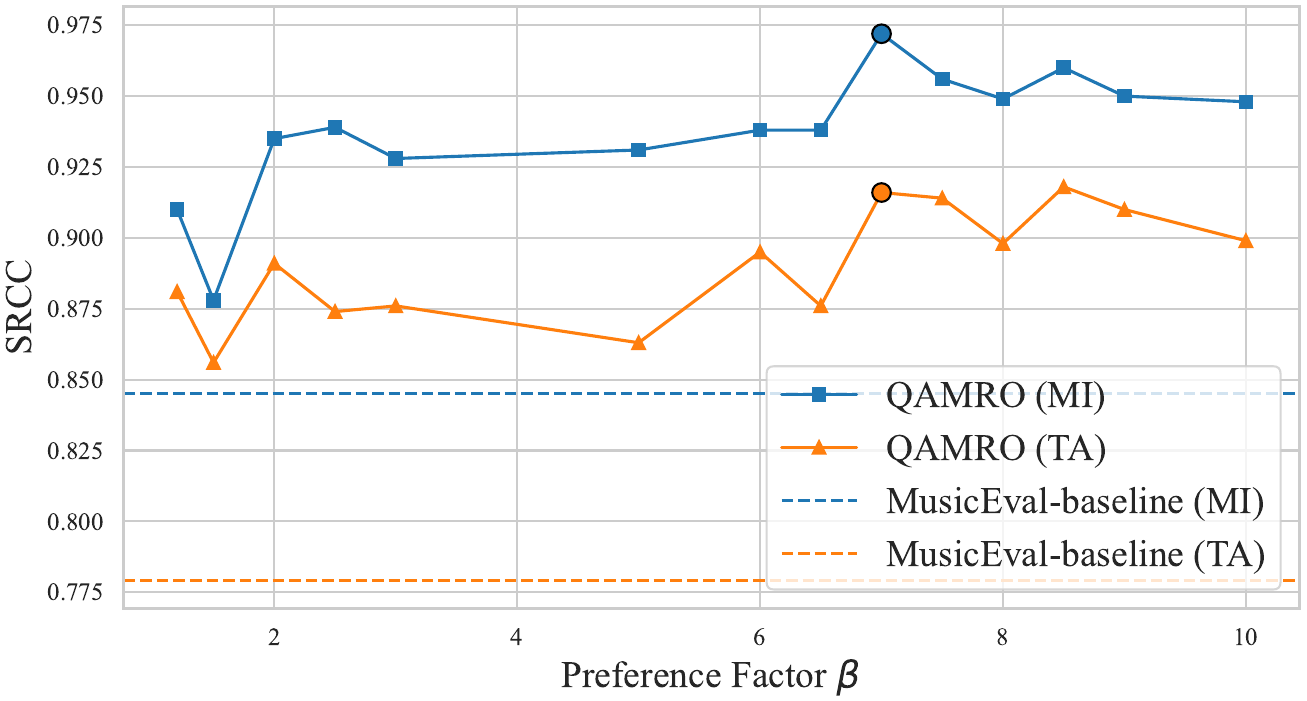}
\vspace{-10pt}
\caption{Effect of the preference factor $\beta$ on the performance of QAMRO in terms of musical impression (MI) and textual alignment (TA) on the MusicEval dataset. Baseline scores are shown for reference.}
\label{fig:beta}
\vspace{-10pt}
\end{figure}

Furthermore, we analyzed the influence of the preference factor $\beta$ in Fig.~\ref{fig:beta}, which controls the emphasis on high-quality samples via the weighting term $q_{ij}$ in Eq.~\ref{eq:weight}.
Performance peaks at $\beta = 7.0$, suggesting that moderate preference weighting improves ranking precision in perceptually sensitive regions.
Larger values slightly degrade performance due to overemphasis on high scores.
These findings highlight the importance of tuning $\beta$ to balance fine-grained discrimination and global consistency.
Notably, all tested $\beta$ values consistently outperform the baseline without quality-aware weighting, demonstrating the effectiveness of our framework.

\subsection{Evaluation Results on AES-Natural}

Table~\ref{tab:track2} presents the results on the AES-Natural dataset, evaluating production quality (PQ), production complexity (PC), content enjoyment (CE), and content usefulness (CU) using SRCC.
QAMRO achieves the best performance across all dimensions, with notable improvements over the Audiobox-Aesthetics baseline \cite{tjandra2025}, particularly on content usefulness.
Moreover, our framework consistently shows effectiveness across different model architectures and various MOS score aspects, demonstrating its robustness and generalizability.

The ablation studies further validate the effectiveness of each component, using the same ablation settings as in Table~\ref{tab:track1}.
The removal of the quality-aware weighting consistently degrades performance across evaluation aspects, with a more pronounced impact observed when the adaptive margin mechanism is also excluded, especially for content usefulness.
These results highlight the importance of both the quality-aware weighting and adaptive margin in capturing nuanced perceptual and semantic aspects of music content.

\section{Conclusion and Future Work}

This study presents QAMRO, a ranking loss with quality-aware adaptive margins for human-aligned audio quality assessment.
Evaluations on the AudioMOS Challenge 2025 benchmarks reveal that QAMRO significantly outperforms strong baselines in correlating with human MOS scores, particularly for high-quality samples, demonstrating the first effective use of ranking loss in this context.
As to future work, we envisage to explore listwise ranking methods apart from our current pairwise strategy, which consider global ranking structures and are anticipated to improve alignment with human judgments across diverse generated audio.

\newpage

\bibliographystyle{IEEEtran}
\bibliography{references}
\end{document}